\documentclass[a4paper,10pt]{article}

\usepackage{subfig}
\usepackage{graphicx}
\usepackage{cite}
\usepackage{bbm}
\usepackage{epsfig,amsmath,amssymb,verbatim,mathrsfs,array,layout,textcomp,amssymb,latexsym}

\newcommand{\beq}{\begin{eqnarray}}
\newcommand{\eeq}{\end{eqnarray}}

\def\beqa{\begin{eqnarray}}
\def\eeqa{\end{eqnarray}}
\newcommand{\no}{\nonumber}
\newcommand{\bv}{\left(\begin{array}{c}}
\newcommand{\ev}{\end{array}\right)}
\newcommand{\bmtwo}{\left(\begin{array}{cc}}
\newcommand{\bmthree}{\left(\begin{array}{ccc}}
\newcommand{\emn}{\end{array}\right)}
\newcommand{\bmtwoc}{\left\{\begin{array}{cc}}
\newcommand{\bmthreec}{\left\{\begin{array}{ccc}}
\newcommand{\emnc}{\end{array}\right\}}
\newcommand{\ba}{\begin{array}}
\newcommand{\ea}{\end{array}}

\newcommand{\half}{\frac{1}{2}}

\newcommand{\ep}{\epsilon_{\text{PQ}}}

\def\lsim{\mathrel{\rlap{\lower4pt\hbox{\hskip1pt$\sim$}}
     \raise1pt\hbox{$<$}}}         
\def\gsim{\mathrel{\rlap{\lower4pt\hbox{\hskip1pt$\sim$}}
     \raise1pt\hbox{$>$}}}         

\begin{document}

\begin{titlepage}

\vskip1.5cm
\begin{center}
  {\Large \bf FN-2HDM: Two Higgs Doublet Models\\
   with Froggatt-Nielsen Symmetry}
\end{center}
\vskip0.2cm

\begin{center}
Avital Dery and Yosef Nir\\
\end{center}
\vskip 8pt

\begin{center}
{\it Department of Particle Physics and Astrophysics\\
Weizmann Institute of Science, Rehovot 76100, Israel} \vspace*{0.3cm}

{\tt   avital.dery,yosef.nir@weizmann.ac.il}
\end{center}

\vglue 0.3truecm

\begin{abstract}
  \vskip 3pt \noindent
  We embed Two Higgs Doublet Models (2HDMs) in the Froggatt Nielsen (FN) framework. We find that the approximate FN symmetry predicts i) approximate Natural Flavor Conservation (NFC) of Types II or IV in the Yukawa sector, and ii) approximate Peccei-Quinn (PQ) symmetry in the scalar sector. We discuss the phenomenological consequences of these features.
\end{abstract}

\end{titlepage}

\section{Introduction}

The Froggatt-Nielsen (FN) mechanism~\cite{Froggatt:1978nt,Leurer:1992wg} applies a symmetry principle to explain the non-trivial structure of the measured flavor observables, characterized by smallness and hierarchy. It postulates that the fermion fields are charged under a symmetry that is explicitly broken by a small parameter. Consequently, the various fermion masses and CKM mixing angles are suppressed by different powers of the symmetry breaking parameter. In this way it solves the Standard Model (SM) flavor puzzle - the question of how the hierarchical structure of the flavor parameters is generated.

The discovery of the Higgs boson sparks renewed interest in the scalar sector of Nature and, in particular, in the possibility that it is non-minimal. Both improved measurements of the Higgs couplings and direct searches for additional scalars provide guidance to the possible structure of such a non-minimal scalar sector. A viable extension of the SM scalar sector is the Two-Higgs-Doublet Model (2HDM), that predicts four additional scalars beyond the SM Higgs boson, some of which may be light (for a recent review, see Ref. \cite{Branco:2011iw}). 2HDMs induce, in general, flavor changing processes that are strongly constrained by experiment. Therefore, a mechanism to control the New Physics (NP) flavor structures is usually applied to them, such as Natural Flavor Conservation (NFC) \cite{Glashow:1976nt,Paschos:1976ay} or Minimal Flavor Violation (MFV) \cite{D'Ambrosio:2002ex}.

In this work we study the effectiveness of the FN mechanism in constraining the flavor structures of 2HDMs. We find that, if the scalar doublets are charged under the FN symmetry, various viable options in model building open up, affecting the flavor structure as well as the scalar spectrum.

The plan of this paper goes as follows. In Section \ref{sec:2HDM} we introduce the 2HDM framework, and present relations among Yukawa matrices that are useful for our purposes. In Section \ref{sec:FN} we impose the approximate FN symmetry on the 2HDM, and obtain the resulting structure of the Yukawa matrices and of the scalar potential. In Section \ref{sec:Exp} we confront the FN-2HDM framework with experimental constraints from electroweak precision tests, collider searches, flavor changing neutral current processes, and electric dipole moment searches. Section \ref{sec:nfcmfv} compares the FN-2HDM to 2HDM frameworks with NFC or with MFV. We summarize our conclusions in Section \ref{sec:con}. Specific examples of FN-2HDMs are presented in an Appendix.

\section{The General Framework}
\label{sec:2HDM}
We set the stage with a general 2HDM. The relevant phenomenology of the model is determined by two sectors: the scalar potential, which in general is given by
\beqa\label{eq:scapot}
V&=& m_1^2|\Phi_1|^2+m_2^2|\Phi_2|^2+\frac12\lambda_1|\Phi_1|^4+\frac12\lambda_2|\Phi_2|^4
+\lambda_3|\Phi_1|^2|\Phi_2|^2+\lambda_4|\Phi_1\sigma_2\Phi_2|^2\no\\
&&+\left[\frac12\lambda_5(\Phi_1^\dagger\Phi_2)^2
+(\Phi_1^\dagger\Phi_2)(m_{12}^2+\lambda_6|\Phi_1|^2+\lambda_7|\Phi_2|^2)+{\rm h.c.}\right],
\eeqa
and the Yukawa interactions, given by
\beq\label{eq:yukawa2}
{\cal L}_Y=\sum_{i=1,2}\left(Q\Phi_i Y^u_i \overline{U}
+Q\tilde\Phi_i Y^d_i \overline{D}+L\tilde\Phi_i Y^e_i \overline{E}\right).
\eeq
Here, $\Phi_1$ and $\Phi_2$ are the two scalar doublets, $Q,\bar{U},\bar{D}$ are the quark doublets, up-singlets and down-singlets respectively, and $L,\bar{E}$ are the lepton doublets and charged lepton singlets, respectively.
The Yukawa matrices, $Y^F_i$, are responsible for the corresponding fermion mass matrices $M^F$, which we represent by the dimensionless matrices $Y^F_M$:
\beq
Y_M^F=\frac{\sqrt2 M^F}{v},\ \ \ (F=u,d,e).
\eeq
We denote by $Y_S^F, S=h,H,A$, the Yukawa couplings of the light CP-even scalar $h$, the heavy CP-even scalar $H$, and the CP-odd scalar $A$. (The Yukawa matrices of the charged Higgs $H^\pm$ are the same as those of $A$.) Each of these matrices is a linear combination of $Y_1$ and $Y_2$:
\beqa\label{eq:defys}
Y_M^F&=&+c_\beta Y_1^F+s_\beta Y_2^F,\nonumber\\
Y_A^F&=&-s_\beta Y_1^F+c_\beta Y_2^F,\nonumber\\
Y_h^F&=&-s_\alpha Y_1^F+c_\alpha Y_2^F,\nonumber\\
Y_H^F&=&+c_\alpha Y_1^F+s_\alpha Y_2^F,
\eeqa
where $c_\beta\equiv\cos\beta$, $s_\beta\equiv\sin\beta$, $\tan\beta\equiv v_2/v_1$, and $\alpha$ is the rotation angle from $({\cal R}e(\phi_1^0),{\cal R}e(\phi_2^0))$ to $(H,h)$. The angle $\beta$ is taken to be in the range $[0,\pi/2]$ while $\alpha\in[-\pi/2,+\pi/2]$.
We can express the Yukawa matrices of the scalars in terms of $Y^F_M$,
\beqa \label{eq:Yrelations}
Y_h^F&=&-\frac{s_\alpha}{c_\beta}Y_M^F+\frac{c_{\alpha-\beta}}{c_\beta}Y_2^F
=\frac{c_\alpha}{s_\beta}Y_M^F-\frac{c_{\alpha-\beta}}{s_\beta}Y_1^F,\no\\
Y_H^F&=&\frac{c_\alpha}{c_\beta}Y_M^F+\frac{s_{\alpha-\beta}}{c_\beta}Y_2^F
=\frac{s_\alpha}{s_\beta}Y_M^F-\frac{s_{\alpha-\beta}}{s_\beta}Y_1^F,\no\\
Y_A^F&=&-\tan\beta Y_M^F+(1/c_\beta)Y_2^F=\cot\beta Y_M^F-(1/s_\beta)Y_1^F.
\eeqa
%

\section{2HDMs within the FN Framework}
\label{sec:FN}
We assume that the smallness and hierarchy exhibited by the pattern of fermion masses and Yukawa couplings is a result of an approximate horizontal symmetry (the FN symmetry) $U(1)_H$. The breaking of this symmetry is characterized by a spurion, $\epsilon_H$, to which we assign FN charge of $H(\epsilon_H)=-1$. The FN charges of the different fields dictate the parametric suppression of couplings in both in Yukawa Lagrangian and the scalar potential. In the following we arrive at two possible features of 2HDMs within the FN framework: approximate Natural Flavor Conservation (NFC), and approximate Peccei-Quinn (PQ) symmetry~\cite{Peccei:1977hh} (with possibly large soft breaking).

\subsection{The Yukawa Matrices}
\label{subsec:YukawaFN}
The entries of the six Yukawa matrices are parametrically suppressed according to
\beqa \label{eq:YukawaSup}
(Y^{u}_{1,2})_{ij}&\sim&\epsilon_H^{|H(Q_i)+H(\bar U_j)+H(\Phi_{1,2})|},\no\\
(Y^{d}_{1,2})_{ij}&\sim&\epsilon_H^{|H(Q_i)+H(\bar D_j)-H(\Phi_{1,2})|},\no\\
(Y^{e}_{1,2})_{ij}&\sim&\epsilon_H^{|H(L_i)+H(\bar E_j)-H(\Phi_{1,2})|}.
\eeqa
We use the symbol $\sim$ to denote that we quote the parametric suppression only, and omit ${\cal O}(1)$ coefficients throughout.
The charges of the scalar doublets, $H(\Phi_1)$ and $H(\Phi_2)$, affect the parametric suppression of the coupling matrices universally. If the scalars carry different $U(1)_H$ charges, $H(\Phi_1)-H(\Phi_2)\neq 0$, then there are charge assignment options for which each sector exhibits one Yukawa matrix whose entries are all suppressed compared to the corresponding entries in the other Yukawa matrix. We refer to this situation as {\textit{approximate NFC}}, since one Yukawa matrix dominates each fermion sector.

More concretely, let us take, without loss of generality, $H(\Phi_1)>H(\Phi_2)$, and treat the case where $H(Q_i)+H(\bar U_j) + H(\Phi_1) > 0$ for all $i,j$. Then, a choice of charges that obeys
\beqa
H(Q_i)+H(\bar U_j)  > -\half (H(\Phi_1)+H(\Phi_2)),
\eeqa
for all $i,j$, leads to
\beqa
Y_1^u \ll  Y_2^u.
\eeqa
Similarly, if, for all $i,j$,  $H(Q_i)+H(\bar D_j) - H(\Phi_2) > 0$ and
\beqa
H(Q_i)+H(\bar D_j) < \half (H(\Phi_1)+H(\Phi_2))
\eeqa
then
\beqa
 Y_1^d \gg  Y_2^d.
\eeqa
A different choice, such that $H(Q_i)+H(\bar D_j) > \half (H(\Phi_1)+H(\Phi_2))$ would lead to $Y_1^d \ll  Y_2^d$. In the same way, in the lepton sector with $H(L_i)+H(\bar E_j) - H(\Phi_2) > 0$, we have either
\beqa
H(L_i)+H(\bar E_j) < \half (H(\Phi_1)+H(\Phi_2)) \qquad \text{leading to} \qquad Y_1^e \gg  Y_2^e,
\eeqa
or
\beqa
H(L_i)+H(\bar E_j) > \half (H(\Phi_1)+H(\Phi_2)) \qquad \text{leading to} \qquad Y_1^e \ll  Y_2^e,
\eeqa
Hence, appropriate choices for the charges can lead to what appears to be approximate NFC of Types I, II, III or IV. However, for Types I and III, where both the up and down sectors couple more strongly to the same scalar doublet, we find that a very large charge difference between the scalars is needed, $|H(\Phi_1)-H(\Phi_2)|>12\left(\frac{\log(0.2)}{\log\epsilon_H}\right)$ (see Appendix~\ref{app:appA} for a detailed explanation), making such models less plausible. Therefore in the following we focus on models with approximate NFC of Types II and IV.

This construction leads to one Yukawa matrix whose entries are all suppressed compared to those of the other, for each fermion sector,
\beq
	(Y^F_a)_{ij}\sim \epsilon_{\text{PQ}} (Y^F_b)_{ij}, \qquad (a,b)=(1,2)\ {\rm or}\ (2,1),
\eeq
where we defined
\beq
	\epsilon_{\text{PQ}}\equiv \epsilon_H ^{|H(\Phi_1)-H(\Phi_2)|}.
\eeq

As we demonstrate below, the clear hierarchy between the two Yukawa matrices may not suffice to ward off large flavor changing couplings of the light Higgs once we rotate to the mass basis. Whether or not the FN mechanism suppresses flavor changing Higgs couplings depends on the parameters of the scalar potential. Specifically, the contribution of one Yukawa matrix to both $Y_M^F$ and $Y_h^F$ remains suppressed compared to that of the other, and approximate NFC is preserved, if $\cot\beta$ is not $\epsilon_{\text{PQ}}$ suppressed.

\subsection{The Scalar Potential}
\label{subsec:ScaPot}
If the scalar doublets, $\Phi_1$ and $\Phi_2$, carry different charges under $U(1)_H$, then certain parameters of the scalar potential are parametrically suppressed. Returning to the potential of Eq.~(\ref{eq:scapot}), the terms on the first line are neutral under $U(1)_H$, while those on the second are not, dictating the parametric suppression of the couplings:
\beq
m_{12}^2/\Lambda^2,\ \lambda_6,\ \lambda_7\sim\ep,\ \ \
\lambda_5\sim\ep^2.
\eeq

The soft breaking by $m_{12}^2$ deserves some discussion. Since $m_{12}^2$ is a dimensionful parameter, there is an ambiguity in the low energy theory as to what scale is involved in determining its suppression. In a full high energy realization of the FN framework, there are at least two scales: The electroweak breaking scale $v$, and a scale $\Lambda$ at which the FN symmetry breaking is communicated to the SM fields. Without an explicit UV completion, it is impossible to determine what is the scale compared to which $m_{12}^2$ is $\ep$-suppressed. In common UV completions, the full theory contains a SM-singlet scalar $S$ whose VEV spontaneously breaks the FN symmetry, and a set of vector like quarks and leptons at a mass scale $\Lambda$, such that $\langle S\rangle/\Lambda=\epsilon_H$. In such models, $m_{12}^2\sim\ep\Lambda^2/(16\pi^2)$. We later find a phenomenological constraint, $\ep\lsim10^{-3}$, which in this specific scenario translates into $\Lambda\gsim$ tens of TeV. On the other hand, since we are interested in the possibility that the FN mechanism allows a full 2HDM at or below the TeV scale, we do not consider the possibility that $m_{12}^2\gg v^2$.

The extremum equations for the vacuum expectation values (VEVs) are given by
\beqa \label{eq:min}
0&=& m_1^2 v_1+\half\lambda_1 v_1^3+\half\lambda_{34} v_1 v_2^2
+\half\Big[\lambda_5 v_1 v_2^2+2m_{12}^2v_2+3\lambda_6 v_1^2v_2+\lambda_7 v_2^3\Big];\no\\
0&=& m_2^2 v_2+\half\lambda_2 v_2^3+\half\lambda_{34} v_1^2 v_2
+\half\Big[\lambda_5 v_1^2 v_2+2m_{12}^2v_1+\lambda_6 v_1^3+3\lambda_7 v_1 v_2^2\Big],
\eeqa
with $\lambda_{34}=\lambda_3+\lambda_4$.
In the FN symmetry limit, $\ep\to0$, the bracketed terms in Eqs.~(\ref{eq:scapot},\ref{eq:min}) vanish and the potential exhibits a global $U(1)_Y\times U(1)_{\rm PQ}$ symmetry. This symmetry is then broken spontaneously, either completely or partially, as the fields acquire non-zero VEVs. We separate the parameter space into two cases:

\subsubsection{\textbf{\emph{Case A: $m_1^2,m_2^2<0$:}}}
	In the case where both mass-squared parameters are negative, both fields acquire non-zero VEVs and $U(1)_{\rm PQ}$ is spontaneously broken. In the symmetry limit, the CP-odd state is the Nambu-Goldstone Boson (NGB), $m_A^2 = 0$.\\
	Turning on the symmetry breaking parameters introduces a non-zero $m_A^2$, parametrically  suppressed.
	The key phenomenological features of Case A can be summarized as follows:
	\begin{itemize}
		\item The spectrum contains a light CP-odd scalar. The parametric suppression of its mass is dictated by
		\beqa
		m_A^2\sim m_{12}^2.
		\eeqa
		\item $\tan\beta$, $\cot\alpha$, $c_{\beta-\alpha}$ are parametrically ${\cal O}(1)$.
		\item The approximate NFC in the Yukawa sector remains unharmed. The diagonal Yukawa couplings of the three scalars deviate from the NFC values at ${\cal O}(\ep)$,
			\beq
			\label{eq:Adiag}
			(Y_h^{d})_{ii} &=& -\frac{s_\alpha}{c_\beta}(Y_M^{d})_{ii}\Big[1+{\cal O}(\ep)\Big],  \qquad (Y_h^u)_{ii} = \frac{c_\alpha}{s_\beta}(Y_M^u)_{ii}\Big[1+{\cal O}(\ep)\Big];  \\ \no
			(Y_H^{d})_{ii} &=& \frac{c_\alpha}{c_\beta}(Y_M^{d})_{ii}\Big[1+{\cal O}(\ep)\Big], \qquad (Y_H^u)_{ii} = \frac{s_\alpha}{s_\beta}(Y_M^u)_{ii}\Big[1+{\cal O}(\ep)\Big];\\ \no
			(Y_A^{d})_{ii} &=& -\tan\beta (Y_M^{d})_{ii}\Big[1+{\cal O}(\ep)\Big]  \qquad (Y_A^u)_{ii} = \cot\beta (Y_M^u)_{ii}\Big[1+{\cal O}(\ep)\Big].
			\eeq
		\item Suppressed off-diagonal Yukawa couplings appear, aligned between the scalars,
			\beq
			\label{eq:Anondiag}
			(Y_h^{d})_{ij} &=& \frac{c_{\beta-\alpha}}{c_\beta}(Y_2^{d})_{ij}, \qquad (Y_h^u)_{ij} = -\frac{c_{\beta-\alpha}}{s_\beta}(Y_1^u)_{ij};\\ \no
			(Y_H^{d})_{ij} &=& \frac{s_{\beta-\alpha}}{c_\beta}(Y_2^{d})_{ij}, \qquad (Y_H^u)_{ij} = -\frac{s_{\beta-\alpha}}{s_\beta}(Y_1^u)_{ij};\\ \no
			(Y_A^{d})_{ij} &=& \frac{1}{c_\beta}(Y_2^{d})_{ij},\qquad (Y_A^u)_{ij} = -\frac{1}{s_\beta}(Y_1^u)_{ij},
			\eeq
			where in our models
			\beq \label{eq:offdiag}
			(Y_2^{d})_{ij} = {\cal O}(\ep)\times (Y_X^d)_{ij}, \quad
			(Y_1^u)_{ij}= {\cal O}(\ep)\times (Y_X^u)_{ij}, \quad
			(Y_X^F)_{ij} \equiv \begin{cases}
				(Y_M^{d})_{jj}\cdot V_{ij}, & \text{for } j>i \\
				(Y_M^{d})_{jj} / V_{ji} & \text{for } j<i
			\end{cases}
			\eeq
			with $V_{ij}$ the CKM or PMNS matrix elements.	
			
	\end{itemize}

\subsubsection{\textbf{\emph{Case B: $m_1^2>0,\, m_2^2<0$}}}
	When only one mass-squared parameter is negative, one VEV is vanishing in the symmetry limit, $v_1=0$, dictating that $c_\beta=c_{\beta-\alpha}=0$.
	A residual $U(1)$ remains unbroken, so there are no (uneaten) NGBs.\\
	Turning on symmetry breaking leads to
	\beq
		c_\beta, s_\alpha, c_{\beta-\alpha}\sim\ep.
	\eeq
	This behavior has non-trivial implications for the flavor structure of the model. We rewrite the entries of the mass and coupling matrices from Eq.~(\ref{eq:defys}) as
	\beqa
	\label{eq:FCNC}
	(Y_M^F)_{ij} &=& s_\beta (Y_2^F)_{ij}\Big(1+\cot\beta \frac{(Y_1^F)_{ij}}{(Y_2^F)_{ij}}\Big)=c_\beta (Y_1^F)_{ij}\Big(1+\tan\beta \frac{(Y_2^F)_{ij}}{(Y_1^F)_{ij}}\Big); \\ \no
	(Y_h^F)_{ij} &=& c_\alpha (Y_2^F)_{ij}\Big(1-\tan\alpha \frac{(Y_1^F)_{ij}}{(Y_2^F)_{ij}}\Big)=-s_\alpha (Y_1^F)_{ij}\Big(1-\cot\alpha \frac{(Y_2^F)_{ij}}{(Y_1^F)_{ij}}\Big).
	\eeqa
	It is then apparent that the approximate NFC persists only if the ${\cal O}(\ep)$ suppression of the ratio of Yukawa entries is not met with the ${\cal O}(\ep^{-1})$ enhancement of $\tan\beta$ or $\cot\alpha$. Otherwise, the mass and coupling matrices receive same order contributions with different order one factors, leading to flavor changing Higgs couplings easily violating existing bounds. This poses a problem for the down and lepton sectors in models of Type II and IV, making Case B non-viable.

In conclusion, considering the parametric structure in the scalar and Yukawa sectors, we are led to the following predictions:
\begin{enumerate}
	\item The Yukawa matrices are close to those of NFC of Types II or IV, with deviations of ${\cal O}(\ep)$.
	\item The CP-odd scalar is a pNGB, with mass dictated by the dimensionful explicit FN-breaking parameter, $m_{12}^2$. This is not a sharp prediction for the order of $m_A$, since the relevant scale compared to which $m_{12}^2$ is suppressed is model dependent, $m_{12}^2\sim \ep \Lambda^2$. In the following we take $m_A \lsim v$.
\end{enumerate}
%

\section{Experimental Constraints}
\label{sec:Exp}
In this section we look at the rough structure of the Yukawa couplings, neglecting ${\cal O}(1)$ factors, and at the scalar spectrum generated by the FN symmetry, and confront them with experimental bounds. For this purpose, the diagonal and non-diagonal Yukawa entries for all scalars are taken as (for the detailed expressions see Eqs.~(\ref{eq:Adiag},\ref{eq:Anondiag})):
\beq
\label{eq:FNstruct}
(Y_S)_{ii} &=& {\cal O}((Y_M)_{ii})\times [1 + {\cal O}(\ep)], \\ \no
(Y_S)_{ij} &=& {\cal O}(\ep)\times \begin{cases}
	(Y_M)_{jj}\cdot V_{ij}, & \text{for } j>i \\
	(Y_M)_{jj} / V_{ji} & \text{for } j<i
\end{cases}.
\eeq
The ${\cal O}(\ep)$ factors are assumed to come with order one phases.
The deviations from NFC, embodied in the flavor non-diagonal couplings and the phases, are constrained by low energy experiments, which set an upper bound on the FN suppression factor, $\ep$. We find that the most stringent bounds come from $\mu\to e\gamma$ and $K^0-\overline{K}{}^0$ mixing: $\ep\lsim10^{-3}$.
Other constraints are relevant for the 2HDM scalar resonances, particularly for the scale of light CP-odd scalar mass, $m_A$. These arise from direct collider searches, from electroweak (EW) precision tests, and from the bound on untagged decays of the Higgs. Interestingly, we find that the most stringent bound on $m_A$ comes from an upper bound on the trilinear coupling, $g_{hAA}$ which, in turn,  implies $m_A\gtrsim m_h/2$.

\subsection{EW Precision Tests}
Mass splittings between the scalar eigenstates are expected to induce a contribution to the $T$ parameter. We find that when $m_{H^\pm}\approx m_H$ is maintained, a large splitting between $m_A$ and $m_{H^\pm}$ is allowed as long as $|c_{\beta-\alpha}|$ is small, as suggested by Higgs data. To quantify this statement, we write the contribution of the additional scalars to the $T$ parameter as~\cite{Baak:2011ze}
\beqa
16\pi m_W^2s_W^2\delta T &=& F(m_{H^\pm}^2,m_A^2) + s_{\beta-\alpha}^2[F(m_{H^\pm}^2,m_H^2)-F(m_A^2,m_H^2)]\\ \no
&+& c_{\beta-\alpha}^2[F(m_{H^\pm}^2,m_h^2)-F(m_A^2,m_h^2)+F(m_W^2,m_H^2)-F(m_W^2,m_h^2)\\ \no
&-& F(m_Z^2,m_H^2)+F(m_Z^2,m_h^2)]+4m_Z^2\bar{B}_0(m_Z^2,m_H^2,m_h^2)-4m_W^2\bar{B}_0(m_W^2,m_H^2,m_h^2),
\eeqa
with
\beqa
F(m_1^2,m_2^2) & \equiv & \frac{m_1^2+m_2^2}{2}-\frac{m_1^2m_2^2}{m_1^2-m_2^2}\log\frac{m_1^2}{m_2^2}, \\ \no
\bar{B}_0(m_1^2,m_2^2,m_3^2) & \equiv & \frac{m_1^2\log m_1^2-m_3^2\log m_3^2}{m_1^2-m_3^2}- \frac{m_1^2\log m_1^2-m_2^2\log m_2^2}{m_1^2-m_2^2}.
\eeqa
For $m_{H^\pm}\approx m_H$, the entire expression is proportional to $c_{\beta-\alpha}^2$, and simplifies to
\beq
\delta T = \frac{c_{\beta-\alpha}^2}{16\pi m_W^2s_W^2} &\times & \left\{ F(m_H^2,m_A^2)+ F(m_H^2,m_h^2)-F(m_A^2,m_h^2)+F(m_W^2,m_H^2)-F(m_W^2,m_h^2)\right. \\ \no
&-& \left. F(m_Z^2,m_H^2)+F(m_Z^2,m_h^2)+4m_Z^2\bar{B}_0(m_Z^2,m_H^2,m_h^2)-4m_W^2\bar{B}_0(m_W^2,m_H^2,m_h^2)\right\}.
\eeq
\begin{figure}[!ht]
	\centering
	\includegraphics[width=3.5in]{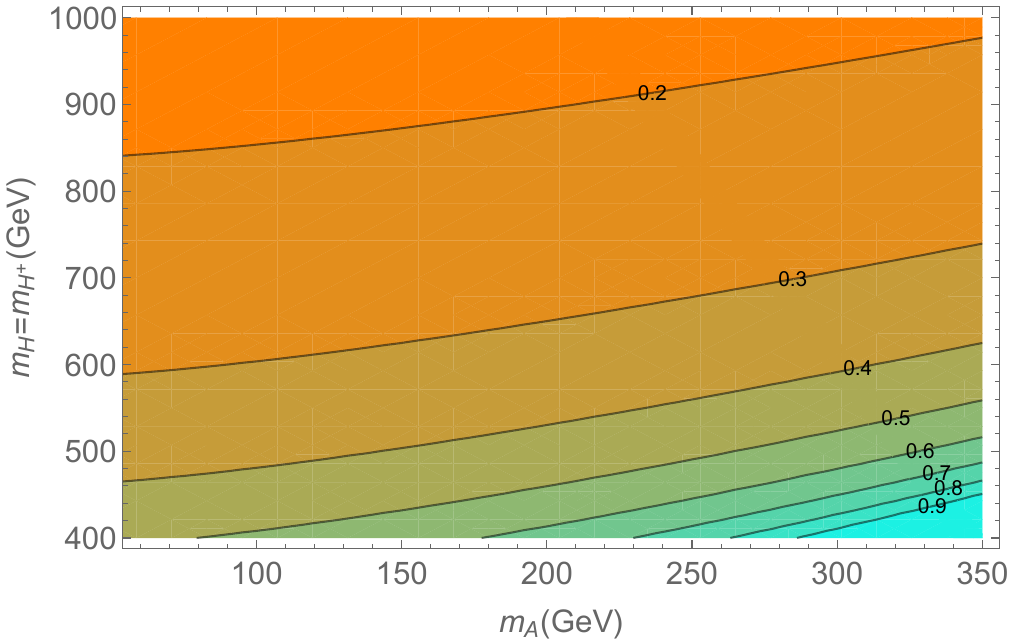}
	\caption{Contours of the upper limit on $|c_{\beta-\alpha}|$, such that the bound on the $T$ parameter, $-0.17\leq T \leq 0.35$ at 95\% C.L.~\cite{Baak:2011ze}, is satisfied.}
	\label{fig:Tparam}
\end{figure}
In Fig.~\ref{fig:Tparam} we plot contours of the upper limit on $|c_{\beta-\alpha}|$ such that the constraint on the $T$ parameter is satisfied, in the plane of $m_H=m_{H^\pm}$ {\it vs.} $m_A$. The charged Higgs mass in models of Types II and IV is constrained by $b\to s\gamma$ measurements to be $\gsim 480\,{\rm GeV}$~\cite{Misiak:2015xwa}. The CP-odd scalar, $A$, may be as light as $\sim m_h/2$, with varying upper limits on $|c_{\beta-\alpha}|$ as $m_H$ is varied, such that $0\leq|c_{\beta-\alpha}|\lsim 0.2-0.6$. The measurement of the Higgs coupling to gauge bosons results in a comparable limit on $|c_{\beta-\alpha}|$~\cite{Aad:2015pla}.

\subsection{Collider Searches}
\label{sec:coll}
Direct searches for the scalar resonances, $A$, $H$ and $H^\pm$, limit the viable spectra of the model. In this regard analyses that are done in the context of Type II and IV NFC with a light CP-odd scalar apply straightforwardly. We consider $m_H\approx m_{H^\pm}\gsim 480\,{\rm GeV}$, to comply with bounds from $b\to s\gamma$~\cite{Misiak:2015xwa}. The remaining relevant constraints involve the light CP-odd scalar, $A$. Since searches for rare meson decays exclude masses below 10 GeV for scalars with ${\cal O}(Y_M)$ diagonal couplings~\cite{Dolan:2014ska}, we consider $m_A\gtrsim 10\,{\rm GeV}$.
\subsubsection{LEP~\cite{Schael:2006cr,Abbiendi:2004ww,Abdallah:2004wy}}
We consider the channels $e^+e^-\to Z^* \to hA$ and $e^+e^-\to Z^*/\gamma^* \to f\bar{f}A$. Pair production of the CP-odd scalar only arises at loop level, through triangle and box diagrams which are suppressed by a loop factor times $m_e^2/m_Z^2$ compared to tree level $hA$ production, making its contribution to the $4f$ final state negligible.

\begin{itemize}
	\item $e^+e^-\to hA$
\end{itemize}
	Taking into account the existing bounds on $|c_{\beta-\alpha}|$~\cite{Aad:2015pla} and BR$(h\to\tau\tau)$~\cite{Aad:2015vsa} and considering $m_A\gtrsim 10\, {\rm GeV}$, searches for $hA\to 2f_1 2f_2$ with $f_1,f_2=b,\tau$ do not set meaningful bounds. For example, the DELPHI and OPAL analyses or Refs.~\cite{Abdallah:2004wy,Abbiendi:2004ww} put upper limits on $\sigma_{hA}\times{\rm BR}(hA\to2f_1 2f_2)$ of ${\cal O}(0.1)$ times the maximal value in the MSSM. These translate into
\beq
c^2_{\beta-\alpha}\times {\rm BR}(h\to f_1f_1)\times {\rm BR}(A\to f_2f_2) < {\cal O}(0.1)
\eeq
which is automatically fulfilled for $|c_{\beta-\alpha}|\lsim 0.6$.

\begin{itemize}	
	\item $e^+e^-\to f\bar{f}A$
\end{itemize}
	Searches for Yukawa production of $A$ in association with $\tau\bar{\tau}$ or $b\bar{b}$~\cite{Abdallah:2004wy} yield bounds on $t_\beta^2$ or $t_\beta^{-2}$ times ${\rm BR}(A\to f\bar{f})$, for limited windows of $m_A$. The bounds are meaningful when the relevant branching ratio is approximately one, which implies, for $t_\beta\gtrsim 1$:
	\beq
	t_\beta \lsim 20- 100 \qquad {\rm for}\,m_A\approx 12\,{\rm GeV}- 50\, {\rm GeV},
	\eeq
	and for Type IV with $t_\beta\lsim 0.7$:
	\beq
	t_\beta \gtrsim 1.6\times 10^{-2} - 3\times 10^{-3} \qquad {\rm for}\,m_A\approx 10\,{\rm GeV}- 30\, {\rm GeV}.
	\eeq
	The excluded regions are plotted in the $t_\beta-m_A$ plane in Fig.~\ref{fig:tbmA}, where we do not consider $t_\beta$ values smaller than ${\cal O}(0.1)$ as they lead to non-perturbative $Y^S_t$.

\subsubsection{LHC}
\begin{itemize}
	
	\item CMS $bbA\to bb\tau\tau$ search~\cite{Khachatryan:2015baw}
\end{itemize}

\begin{figure}[!ht]
	\centering
	\subfloat[Type II]{\includegraphics[width=0.42\textwidth]{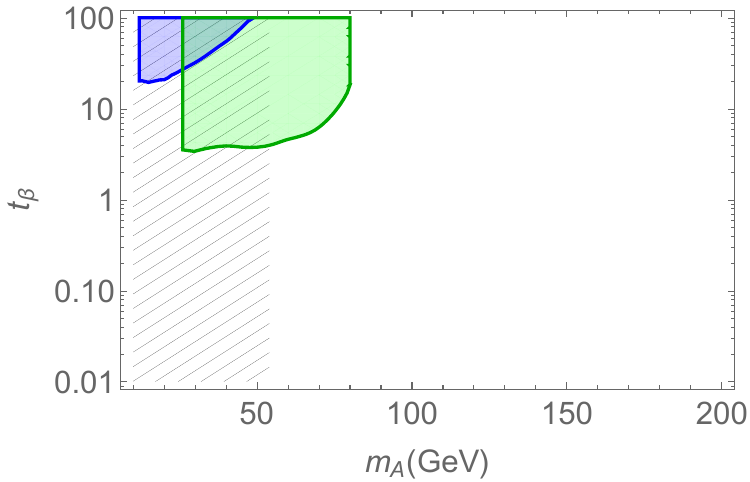}}
	\subfloat[Type IV]{\includegraphics[width=0.67\textwidth]{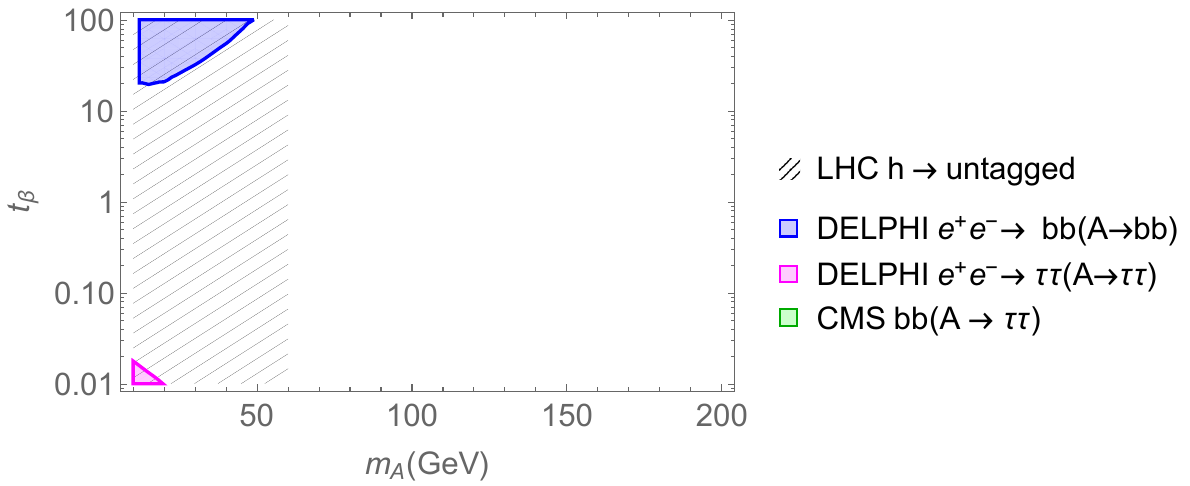}}
	\caption{Excluded regions in the $t_\beta-m_A$ plane for (a) Type II and (b) Type IV models, at 95\% C.L. from various measurements at LEP and the LHC.}
	\label{fig:tbmA}
\end{figure}

	A search for associated production of a scalar along with two b-jets was performed for $25\,{\rm GeV}\leq m_A \leq 80\,{\rm GeV}$, in the $\tau\bar{\tau}$ decay channel. The analysis puts bounds on models of Type II with $t_\beta>1$, for which we use ${\rm BR}(A\to \tau\tau)\approx \frac{m_\tau^2}{3m_b^2}\frac{\beta_\tau^3}{\beta_b^3}$ (where $\beta_x\equiv (1-{4m_x^2}/{m_A^2})^{1/2}$). We can estimate the production cross section using MG5@NLO~\cite{Alwall:2014hca}:
	\beq
	\sigma_{\text{tree}}^8(b\bar{b}A)&=& 16\,{\rm pb}\times t_\beta^2 \qquad \text{for} \,\, m_A=25\,{\rm GeV}\\
	\sigma_{\text{tree}}^8(b\bar{b}A)&=& 0.9\,{\rm pb}\times t_\beta^2 \qquad \text{for} \,\, m_A=80\,{\rm GeV}\\
	\eeq
	The induced limit on $t_\beta$, depicted in Fig.~\ref{fig:tbmA}, is of the order of
	\beq
	t_\beta\lsim 3 - 17.5, \qquad \text{for}\,\, m_A= 25\, {\rm GeV} - 80\,{\rm GeV}.
	\eeq
	For Type IV this search sets no significant limit.

\begin{figure}[!ht]
		\centering
		\subfloat[Type II]{\includegraphics[width=0.5\textwidth]{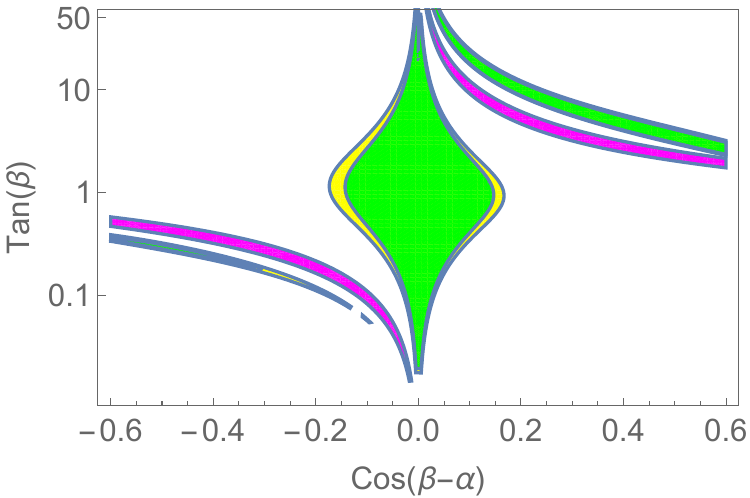}}
		\subfloat[Type IV]{\includegraphics[width=0.5\textwidth]{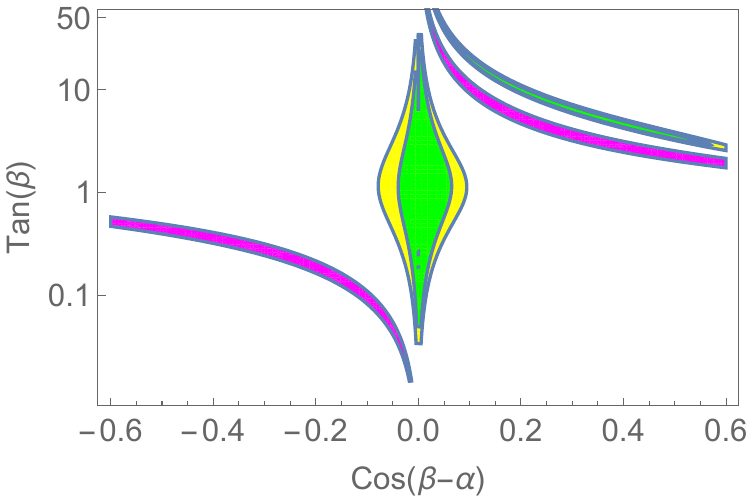}}
		\caption{Viable regions in the $t_\beta-c_{\beta-\alpha}$ plane for (a) Type II and (b) Type IV models, at one (green) and two (yellow) sigma.  The region satisfying the upper bound on ${\rm BR}(h\to AA)$ for $m_A=30\,{\rm GeV}$ is given in magenta.}
		\label{fig:ghAA}
	\end{figure}

\begin{itemize}	
	\item Untagged Higgs Decays
\end{itemize}	
	For $m_A\lsim \half m_h$, $h\to AA$ proceeds through the trilinear coupling, $g_{hAA}$. This dimensionful coupling is naively of order the weak scale, making the channel $h\to AA$ the dominant decay mode for any value of $m_A$ in this range, in tension with the existing bound~\cite{Bernon:2014vta},
	\beq \label{eq:untagged}
	{\rm BR}(h\to \text{untagged})\lsim 0.3.
	\eeq
	Assuming no other non-SM decay modes of the Higgs exist, a rough bound can be obtained by requiring that  $\Gamma(h\to AA)$ is smaller than $\Gamma^{\rm SM}(h\to b\bar{b})$ at tree level:
	\beq\label{eq:naiveg}
	\frac{\beta_A}{12\beta_b^3}\left(\frac{g_{hAA}v}{m_h m_b}\right)^2\lsim 1,
	\eeq
	with $\beta_A\equiv (1-{4m_A^2}/{m_h^2})^{1/2}$. This translates into the bound
	\beq
	g_{hAA} \lsim \frac{9\, {\rm GeV}}{(1-4m_A^2/m_h^2)^{1/4}}.
	\eeq
	For a  general 2HDM, the $g_{hAA}$ coupling is related to the mass-squared parameters by~\cite{Gunion:2002zf}
	\beq \label{eq:ghAA}
	g_{hAA}=-\frac{1}{v}\left[s_{\beta-\alpha}\left(m_h^2+2m_A^2-2\frac{m_{12}^2}{s_\beta c_\beta}\right)-c_{\beta-\alpha}(t_\beta-t_\beta^{-1})\left(m_h^2+\frac{1}{c_{2\beta}}(m_A^2-2c_\beta^{2}\frac{m_{12}^2}{s_\beta c_\beta})\right)\right].
	\eeq
	The FN structure dictates that $\frac{m_{12}^2}{s_\beta c_\beta}=m_A^2+{\cal O}(\ep v^2)$, leading to
	\beq\label{eq:ghAA_FN}
	 g_{hAA}=-\frac{m_h^2}{v}\left[s_{\beta-\alpha}-c_{\beta-\alpha}(t_\beta-t_\beta^{-1})\left(1-\frac{m_A^2}{m_h^2}\right)+{\cal O}(\ep)\right].
	\eeq
	The bracketed expression in Eq.~(\ref{eq:ghAA_FN}) bears some resemblance to the expressions for the normalized Yukawa couplings of Type II/IV NFC:
	\beq
	\kappa_u\equiv y_h^u/y^u_{\rm SM} &=& s_{\beta-\alpha} + c_{\beta-\alpha}t_\beta^{-1},\\ \no
	\kappa_d\equiv y_h^d/y^d_{\rm SM} &=& s_{\beta-\alpha} - c_{\beta-\alpha}t_\beta.
	\eeq
	Higgs coupling measurements constrain these normalized couplings to be not far from unity, in absolute value, while the expression in Eq.~(\ref{eq:ghAA_FN}) needs to be small in order to keep the branching ratio to $AA$ from dominating.
	The tension created between the constraints from the Type II/IV 2HDM global Higgs fit and the added constraint on untagged decays is demonstrated in Fig.~\ref{fig:ghAA}, where the constraint from BR$(h\to AA)$ is plotted in magenta, for $m_A= 30 \,{\rm GeV}$.
The top-right magenta strand follows $c_{\beta-\alpha} t_\beta \approx 1$, while the close-by green ``peninsula" region of the Higgs fit follows $c_{\beta-\alpha}t_\beta \approx 2$. In a similar manner, the bottom left Higgs fit and magenta regions follow $c_{\beta-\alpha}t_\beta^{-1}\approx -2$ and $c_{\beta-\alpha}t_\beta^{-1}\approx -1$, respectively. The width of the  $g_{hAA}$ (magenta) regions varies with $m_A$ and becomes unbound as $m_A$ approaches $m_h/2$.
	
This constraint excludes $m_A$ below $m_h/2$ except for the range approaching the threshold (as the phase space goes to zero). More precisely, we get $m_A\gtrsim 54\, {\rm GeV}$ for Type II approximate NFC, and $m_A\gtrsim 60 \, {\rm GeV}$ for Type IV approximate NFC.

Fig.~\ref{fig:tbmA} summarizes the constraints on the mass of the CP-odd scalar for different values of $t_\beta$. We note that $m_A \lsim m_h/2$ is excluded, while for $m_A\gsim m_h/2$ the parameter space is very weakly constrained.
%

\subsection{Bounds on Deviations from NFC}
Limits on processes that are sensitive to flavor off-diagonal couplings or to non-zero phases constrain the FN suppression factor, $\ep$. The most relevant flavor changing processes are those that involve $Y^S_{e\mu}$, $Y^S_{uc}$, $Y^S_{ds}$ and $Y^S_{ut}$. The bound on the electric dipole moment (EDM) of the electron constrains the deviations from NFC via the imaginary parts of $Y^S_t$ and $Y^S_e$.
\subsubsection{$\mu\to e\gamma$}
\begin{figure}[!ht]
	\centering
	\includegraphics[width=4.5in]{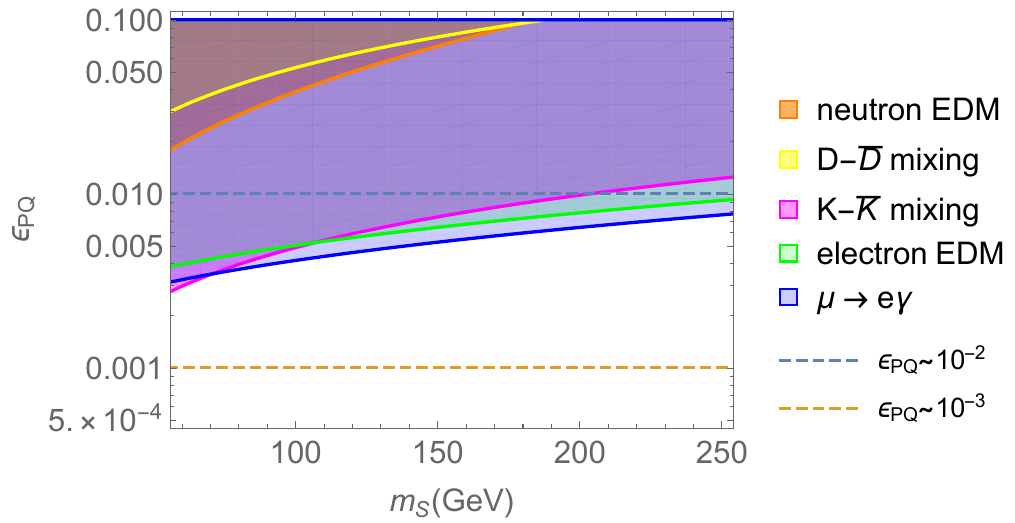}
	\caption{The excluded regions in the $\ep-m_S$ plane from flavor changing neutral current processes and from EDMs.}
	\label{fig:Exclude}
\end{figure}

Following Ref.~\cite{Chang:1993kw}, we parameterize the leading contributions to the process $\mu\to e \gamma$ at one- and two-loops using the reduced amplitudes $A_{L,R}$:
\beqa
\label{eq:mueWilson}
	A_{L,R}^{{\rm 1\,loop},(h,H)} &=& -\frac{1}{24m_S^2}Y^{h,H}_\mu Y^{h,H}_{e\mu,\mu e}\left(4+3\log\frac{m_\mu^2}{m_{h,H}^2}\right)\\ \no
	A_{L,R}^{{\rm 1\,loop},A} &=& -\frac{1}{24m_S^2}Y^A_\mu Y^A_{e\mu,\mu e}\left(5+3\log\frac{m_\mu^2}{m_A^2}\right)\\ \no
	A_{L,R}^{\rm 2\, loop} &=& -\frac{\alpha}{3\pi}\frac{Y^S_t Y^S_{e\mu,\mu e}}{m_\mu m_t}f(z_{tS}),
\eeqa
where $S=h,H$, $z_{tS}=m_t^2/m_S^2$. For the CP-odd scalar $A$, $f(z_{tS})$ needs to be replaced by $g(z_{tS})$. The loop functions read
\beq \label{eq:loopfuncs}
	f(z)&=&\frac{1}{2}z\int_0^1 dx \frac{1-2x(1-x)}{x(1-x)-z}\log\frac{x(1-x)}{z},\\ \no
	g(z)&=&\frac{1}{2}z\int_0^1 dx \frac{1}{x(1-x)-z}\log\frac{x(1-x)}{z}.
\eeq
The rate is the given by
\beq
	\Gamma(\mu\to e \gamma) = \frac{\alpha m_\mu^5}{64\pi^4}(|A_L|^2+|A_R|^2).
\eeq
Substituting the FN structure of Eq.~(\ref{eq:FNstruct}) for the various Yukawa couplings and imposing the MEG bound~\cite{TheMEG:2016wtm}, we arrive at the constraint on the order of parametric suppression as a function of the scalar mass. Fig.~\ref{fig:Exclude} shows the obtained excluded region. In the plot, we use the loop function suitable for the CP-odd scalar as it gives slightly stronger bounds.

As the bound becomes more stringent with smaller mass, we focus on the light CP-odd scalar, which can be as light as ${\cal O}(m_h/2)$ (see Section~\ref{sec:coll}). For $m_A\approx m_h/2$, $\ep\lsim 3\times 10^{-3}$ is required.

\subsubsection{$K^0-\overline{K}{}^0$ and $D^0-\overline{D}{}^0$ Mixing}
Tree level processes involving non-diagonal Yukawa couplings contribute to $K^0-\overline{K}{}^0$ and $D^0-\overline{D}{}^0$ mixing. Following Refs.~\cite{Harnik:2012pb,Bona:2007vi,Isidori:2010kg}, we write the leading contributions to the Wilson coefficients inducing $D^0-\overline{D}{}^0$ and $K^0-\overline{K}{}^0$ mixing as
\beq
	C_2^{uc}=-\frac{(Y^S_{uc})^2}{8m_S^2}, \qquad C_4^{ds}=-\frac{{Y^S_{ds}}^*Y^S_{sd}}{4m_S^2}.
\eeq
From this we infer the constraint on the $\ep$ as a function of $m_S$, shown in Fig.~\ref{fig:Exclude}. For $m_A\approx m_h/2$, the bound reads $\ep\lsim 3\times 10^{-3}$.

\subsubsection{Neutron EDM}
The bound on the electric dipole moment (EDM) of the neutron \cite{Afach:2015sja} implies a bound on the flavor violating couplings $Y_{ut}$ and $Y_{tu}$, through the one-loop contribution to the up quark EDM~\cite{Harnik:2012pb,Pospelov:2005pr},
\beq
	\frac{d_u}{e}\approx -\frac{{\rm Im}(Y^S_{ut}Y^S_{tu})}{16\pi^2}\frac{m_t}{m_S^2}\left(2\log\frac{m_S^2}{m_t^2}-3\right).
\eeq
Calculating the subsequent contribution to the neutron EDM results in a bound on $\ep$ shown in Fig.~\ref{fig:Exclude}. For $m_A\approx m_h/2$, the bound reads $\ep\lsim 3\times 10^{-2}$

\subsubsection{Electron EDM}
The deviations from NFC in the diagonal Yukawa couplings are also ${\cal O}(\ep)$ (see Eq.~(\ref{eq:FNstruct})), and in general are accompanied by ${\cal O}(1)$ phases. We therefore derive bounds on the order of these deviations from the bound on the electron EDM \cite{Baron:2013eja} which is sensitive to the phases in $Y^S_t$ and $Y^S_e$. Following Ref.~\cite{Brod:2013cka}, we write the contribution of the Barr-Zee type diagram to the electron EDM as
\beq
	\frac{d_e}{e}=\frac{8}{3}\frac{\alpha}{(4\pi)^3}\frac{1}{m_t}[{\rm Re}(Y^S_e){\rm Im}(Y^S_t)f(x_{tS})+{\rm Im}(Y^S_e){\rm Re}(Y^S_t)g(x_{tS})],
\eeq
with $x_{tS}\equiv m_t^2/m_S^2$. The loop functions are given in Eq.~(\ref{eq:loopfuncs}).
We use ${\rm Re}(Y^S_f)={\cal O} (Y^M_f)$ and ${\rm Im}(Y^S_f)= {\cal O} (\ep\cdot Y^M_{f})$ to estimate the bound on the suppression factor as a function of $m_S$, as shown in Fig.~\ref{fig:Exclude}. For $m_A\approx m_h/2$, we get $\ep\lsim 4\times 10^{-3}$.

To summarize the impact of the available experimental input on the parameters of the FN-2HDM model, we find the following:
\begin{enumerate}
	\item The CP-odd scalar mass is constrained to be $m_A\gtrsim m_h/2$.
	\item The charged Higgs mass is constrained to be $m_{H^\pm}\gtrsim 480 {\rm GeV}$.
 \item The CP-even heavy Higgs is quasi-degenerate with the charged Higgs, $m_H\approx m_{H^{\pm}}$.
\item The Yukawa couplings of all scalars are close to those of type II or IV NFC, with ${\cal O}(\ep)$ deviations in diagonal couplings, and the appearance of ${\cal O}(\ep Y_X^S)$ off-diagonal couplings (see Eq.~(\ref{eq:offdiag})). In order to comply with current bounds, $\ep < {\cal O}(10^{-3})$ is required.
\end{enumerate}

\section{Comparison to Other Flavor Frameworks}
\label{sec:nfcmfv}
The FN mechanism was suggested to both explain the smallness and hierarchy in the SM flavor parameters and solve the New Physics flavor puzzle, {\it i.e.} allow new physics (such as the 2HDM) at the TeV scale without violating flavor related bounds. There are additional mechanisms that were suggested to solve the new physics flavor puzzle, such as Natural Flavor Conservation (NFC) and Minimal Flavor Violation (MFV). In this section we compare the predictions of the FN-2HDM model, to 2HDM models with either NFC or MFV.

On the qualitative level, we make the following observations. The SM predicts that the Yukawa couplings have the features of proportionality, $y_i/y_j=m_i/m_j$, and diagonality, $y_{ij}=0$ for $i\neq j$. NFC maintains these two features, though the factor of proportionality in $y_i/m_i$ can be different from the SM prediction of $\sqrt2/v$. MFV gives deviations from proportionality and diagonality that are flavor dependent, with larger deviations for heavier generations. FN gives deviations from proportionality that are all of ${\cal O}(\ep)$ and flavor dependent deviations from diagonality. These qualitative features are demonstrated in Table~\ref{tab:nfcmfvfn}. A more quantitative discussion is given in the next two subsections.

\begin{table}[t]
	\begin{center}
		\begin{tabular}{cccc} \hline\hline
			\rule{0pt}{1.2em}%
			Yukawa Coupling &  NFC  & MFV & FN \\[2pt] \hline\hline
			\rule{0pt}{1.2em}%
$\frac{y_\mu/y_\tau}{m_\mu/m_\tau}$ & $1$ & $1+{\cal O}(y_\tau^2)$ & $1+{\cal O}(\ep)$ \\
$y_{\mu\tau}$ & $0$ & $0$ & ${\cal O}(y_\tau U_{\mu3}\ep)$ \\
$\frac{y_c/y_t}{m_c/m_t}$ & $1$ & $1+{\cal O}(y_t^2)$ & $1+{\cal O}(\ep)$ \\
$y_{ct}$ & $0$ & ${\cal O}(y_t y_b^2 V_{cb}V_{tb}^*)$ & ${\cal O}(y_t V_{cb}\ep)$ \\
			\hline\hline
		\end{tabular}
	\end{center}
	\caption{Examples of deviations from proportionality and diagonality in the NFC, MFV and FN frameworks. (For leptonic MFV we assume here that $Y^e$ is the only spurion.)}\label{tab:nfcmfvfn}
\end{table}

\subsection{NFC-2HDM}
The implementation of a 2HDM within a FN symmetry framework predicts approximate NFC of Types II or IV. Deviations from NFC in the magnitude of diagonal Yukawa couplings are at most of ${\cal O}(10^{-3})$, which is unobservably small.
The existence of phases in diagonal couplings, and of off-diagonal Yukawa couplings in all fermion sectors marks a qualitative departure from NFC predictions. The deviations from NFC are all linked, up to ${\cal O}(1)$ factors, by a common parametric suppression, which provides a rough prediction relating possible future deviations from the SM in experiments.

Table~\ref{tab:nonNFC} presents the upper bounds on the deviations from NFC in various Yukawa couplings, along with the current experimental sensitivity.

\begin{table}[t]
	\begin{center}
		\begin{tabular}{ccc} \hline\hline
			\rule{0pt}{1.2em}%
			Yukawa Coupling &  Order of Magnitude  & Current Sensitivity\\[2pt] \hline\hline
			\rule{0pt}{1.2em}%
			$|Y_{tc}|$ & ${\cal O}(10^{-4}\frac{\ep}{10^{-3}})$ & ${\cal O}(10^{-1})$~\cite{Aad:2015pja} \\
			$|Y_{uc}|$ & ${\cal O}(10^{-6}\frac{\ep}{10^{-3}})$ & ${\cal O}(10^{-5})$~\cite{Bona:2007vi}\\
			$|Y_{ds}Y_{sd}|$ & ${\cal O}(10^{-14}\frac{\ep}{10^{-3}})$ & ${\cal O}(10^{-13})$~\cite{Bona:2007vi}\\
			$|Y_{bs}|$ & ${\cal O}(10^{-5}\frac{\ep}{10^{-3}})$ & ${\cal O}(10^{-3})$~\cite{Bona:2007vi}\\
			$|Y_{e\tau}|$ & ${\cal O}(10^{-6}\frac{\ep}{10^{-3}})$ & ${\cal O}(10^{-3})$~\cite{Aad:2016blu,Khachatryan:2016rke}\\	
			$|Y_{\mu\tau}|$ & ${\cal O}(10^{-5}\frac{\ep}{10^{-3}})$ & ${\cal O}(10^{-2})$~\cite{Aad:2016blu,Khachatryan:2015kon}\\	
			$|Y_{e\mu}|$ & ${\cal O}(10^{-6}\frac{\ep}{10^{-3}})$ & ${\cal O}(10^{-6})$~\cite{TheMEG:2016wtm}\\	
			${\rm Im}(Y_t)$ & ${\cal O}(10^{-3}\frac{\ep}{10^{-3}})$& ${\cal O}(10^{-2})$~\cite{Baron:2013eja}\\
			\hline\hline
		\end{tabular}
	\end{center}
	\caption{The order of non-NFC Yukawa couplings in the FN-2HDM, and the current experimental sensitivity to these couplings.}\label{tab:nonNFC}
\end{table}
\subsection{MFV-2HDM~\cite{Dery:2013aba}}
The FN symmetry predicts proportionality, $Y^F_S\propto Y_M^F$,, with deviations of order $\ep$. The deviations are of the same order for all proportionality relations, for example,
\beq
	\frac{Y^S_\mu/m_\mu}{Y^S_\tau/m_\tau} = \frac{Y^S_c/m_c}{Y^S_t/m_t}=1+{\cal O}(\ep).
\eeq
Approximate proportionality is also a feature of MFV, in which case the deviations are flavor-dependent. For $(Y^S_\mu/Y^S_\tau)/(m_\mu/m_\tau)$ the deviation from unity would be of ${\cal O}(m_\tau^2/v^2)$.
Both frameworks exhibit relations between off-diagonal couplings involving the masses and mixing angles, with some qualitatively different predictions. The overall trend is that the hierarchy between off-diagonal elements involving third generation fermions and those that do not involve the third generation is stronger for MFV than for FN. Taking into account the experimental sensitivity, which is in general poorer when third generation fermions are involved, this hints at a possible way to distinguish between the two frameworks. For example (incorporating neutrino-related spurions and assuming NH), MFV predicts
\beq
	(Y^S_{e\tau})_{\rm MFV} \approx \frac{U_{31}^*U_{33}}{U_{31}^*U_{32}}\frac{m_\tau}{m_\mu}\times(Y^S_{e\mu})_{\rm MFV} \sim 10\times(Y^S_{e\mu})_{\rm MFV},
\eeq
while FN predicts
\beq
(Y^S_{e\tau})_{\rm FN} = {\cal O}\left(\frac{U_{13}}{U_{12}}\frac{m_\tau}{m_\mu}\right)(Y^S_{e\mu})_{\rm FN} \sim (Y^S_{e\mu})_{\rm FN}.
\eeq
This implies that $Y^S_{e\tau}/Y^S_{e\mu}$ is suppressed in FN in comparison with MFV predictions. Bearing in mind that the experimental sensitivity for measuring $Y^S_{e\mu}$ is ${\cal O}(10^4)$ times that of $Y^S_{e\tau}$, we reach the conclusion that if $\tau\to e\gamma$ is observed while $\mu\to e \gamma$ is not, FN will be strongly disfavored.

Similarly, in the quark sector, we compare the ratios governing $K^0$-oscillations vs. $B^0$ and $B_s$-oscillations, under MFV and FN. We note that for MFV the dominant contributions are of the form $Y_{ij}^2$, while for FN, $(Y_{ij}Y_{ji})$ contributions turn out to be greater. The relevant expressions are then
\beqa
	({Y_{sb}^S}^2)_{\rm MFV} &=& {\cal O}\left(\left(\frac{V_{ts}V_{tb}}{V_{td}V_{ts}}\right)^2\frac{y_b^2}{y_s^2}\right)\times ({Y^S_{ds}}^2)_{\rm MFV} \sim 10^{8}\times({Y^S_{ds}}^2)_{\rm MFV},\no\\
	({Y_{sb}^S}^*Y^S_{bs})_{\rm FN} &=& {\cal O}\left(\frac{y_b}{y_d}\right)\times (Y^S_{ds}Y^S_{sd})_{\rm FN} \sim 10^{3}\times (Y^S_{ds}Y^S_{sd})_{\rm FN},
\eeqa
and
\beqa
	({Y^S_{db}}^2)_{\rm MFV} &=& {\cal O}\left(\left(\frac{V_{td}V_{tb}}{V_{td}V_{ts}}\right)^2\frac{y_b^2}{y_s^2}\right)\times ({Y_{ds}^S}^2)_{\rm MFV} \sim 10^{7} \times ({Y_{ds}^S}^2)_{\rm MFV},\no\\
	(Y^S_{db}Y^S_{bd})_{\rm FN} &=& {\cal O}\left(\frac{y_s}{y_b}\right)\times ({Y_{ds}^S}^*Y^S_{sd})_{\rm FN} \sim 10^{2} \times ({Y_{ds}^S}^*Y^S_{sd})_{\rm FN} .
\eeq
Considering the experimental sensitivity, which is currently ${\cal O}(10^4)-{\cal O}(10^6)$ greater for observables related to $K^0$-oscillations versus those of $B_d^0$- and $B_s^0$-oscillations, this suggests that if deviations from the SM predictions are measured in $B_s^0$ or $B_d^0$ systems, but not in the $K^0$ system, then FN will be disfavored.

\section{Discussion and Conclusions}
\label{sec:con}
We considered FN-2HDM models. These are two Higgs doublet models subject to an approximate Froggatt-Nielsen symmetry, where the two scalar doublets carry different FN charges. Our main conclusions and findings are the following:
\begin{itemize}
\item For FN-2HDM scenarios in which only one scalar acquires a VEV in the FN symmetry limit, the FN framework is unable to prevent large flavor-changing rates which are in contradiction with experiments. Thus, in viable FN-2HDM, the scalar potential symmetries are broken completely by the VEVs.
\item The FN structure of the Yukawa matrices induces \textit{approximate NFC} of types II or IV. Models of types I and III require very  large charge differences, and are less plausible.
\item The FN structure of the scalar potential induces \textit{approximate PQ symmetry}, with possibly large soft breaking.
\item The viable models predict departure from NFC in the form of new CP violating phases, off-diagonal Yukawa couplings and deviations in the diagonal couplings.
\item The departure from NFC predictions is governed by the PQ symmetry breaking parameter, $\ep$, which we find to be constrained by $\ep\lsim 10^{-3}$ from low energy experiments.
\item When set against the predictions of models of MFV we find that within FN the hierarchy between flavor changing couplings involving the light generations compared to those involving third generation fermions is softened. As a result (together with the experimental inclination to better measure processes of the light fermions), if deviations from SM predictions are measured in flavor-changing processes involving the third generation but not in  the corresponding processes involving light fermions, then FN-2HDMs will be disfavored.
\end{itemize}

\vspace{1cm}
\begin{center}

{\bf Acknowledgements}
\end{center}

We thank Aielet Efrati, Kfir Blum and Daniel Aloni for many helpful discussions.
YN is the Amos de-Shalit chair of theoretical physics.
YN is supported by grants from the I-CORE program of the Planning and Budgeting Committee and the Israel Science Foundation (grant number 1937/12), the Israel Science Foundation (grant number 394/16), the United States-Israel Binational Science Foundation (BSF), Jerusalem, Israel (grant number 2014230), and the Minerva Foundation.

.

\appendix
\section{Examples for Model Building}
\label{app:appA}
\subsection{Approximate NFC - Types II and IV}
We present here a concrete example for the setup of a Type II approximate NFC. We take $\epsilon_H$ of order the Cabbibo angle $\sim 0.2$ (the results can be easily generalized for any value of the symmetry breaking parameter). We take the following charges under $U(1)_H$:
\beqa\label{eq:chargeii}
&&Q_{1,2,3}(4,3,1),\ \ \ \bar U_{1,2,3}(4,1,0),\ \ \ \bar D_{1,2,3}(4,3,3),\no\\
&&L_{1,2,3}(4,4,4),\ \ \ \bar E_{1,2,3}(6,3,1),\no\\
&&\Phi_{1,2}(+2,-1).
\eeqa
The parametric suppression of the matrix elements is given by
\beqa\label{eq:masmatii}
&&Y^u_2\sim\begin{pmatrix}
	\epsilon_H^{7}&\epsilon_H^{4}&\epsilon_H^3\\
	\epsilon_H^{6}&\epsilon_H^{3}&\epsilon_H^2\\
	\epsilon_H^{4}&\epsilon_H^1&\epsilon_H^0\\
\end{pmatrix},\ \ \
(Y^u_1)_{ij}\sim\epsilon_H^3(Y^u_2)_{ij},\no\\
&&Y^d_1\sim
\begin{pmatrix}
	\epsilon_H^{6}&\epsilon_H^{5}&\epsilon_H^{5}\\
	\epsilon_H^{5}&\epsilon_H^{4}&\epsilon_H^{4}\\
	\epsilon_H^{3}&\epsilon_H^{2}&\epsilon_H^{2}\\
\end{pmatrix},\ \ \
(Y^d_2)_{ij}\sim\epsilon_H^3(Y^d_1)_{ij},\no\\
&&Y^e_1\sim
\begin{pmatrix}
	\epsilon_H^{8}&\epsilon_H^{5}&\epsilon_H^{3}\\
	\epsilon_H^{8}&\epsilon_H^{5}&\epsilon_H^{3}\\
	\epsilon_H^{8}&\epsilon_H^{5}&\epsilon_H^{3}\\
\end{pmatrix},\ \ \
(Y^e_2)_{ij}\sim\epsilon_H^3(Y^e_1)_{ij},
\eeqa
thus correctly reconstructing the known pattern of masses and mixing angles.
In a similar way, we can construct a model of approximate NFC of Type IV by changing only the charge assignments of the lepton fields,
\beq
L_{1,2,3}(-3,-3,-3),\ \ \ \bar E_{1,2,3}(-6,-3,-1).
\eeq
We obtain, instead of the lepton Yukawa matrices of Eq. (\ref{eq:masmatii}),
\beq\label{eq:masmativ}
Y^e_2\sim
\begin{pmatrix}
	\epsilon_H^{8}&\epsilon_H^{5}&\epsilon_H^{3}\\
	\epsilon_H^{8}&\epsilon_H^{5}&\epsilon_H^{3}\\
	\epsilon_H^{8}&\epsilon_H^{5}&\epsilon_H^{3}\\
\end{pmatrix},\ \ \
(Y^e_1)_{ij}\sim\epsilon_H^3(Y^e_2)_{ij}.
\eeqa
%

\subsection{Approximate NFC - Type I and III}
Constructing approximate NFC of Types I and III involves large charges and a suppression that is not universal in the down sector. This stems from the different sign that comes with the scalar charge in the up sector (due to the different Hypercharge). As in Eq.~\ref{eq:YukawaSup}, we have
\beq
	(Y^{u}_{1,2})_{ij}&\sim&\epsilon_H^{|H(Q_i)+H(\bar U_j)+H(\Phi_{1,2})|},	
\eeq
while
\beq
	\no\\
	(Y^{d}_{1,2})_{ij}&\sim&\epsilon_H^{|H(Q_i)+H(\bar D_j)-H(\Phi_{1,2})|}.
\eeq
We demonstrate the difficulty that arises with concrete examples, in the next subsections.

Let us assume, without loss of generality, that $H(\Phi_1) > H(\Phi_2)$.
The requirement in Types I and III, that the down sector couples more strongly to $\Phi_2$, like the up sector, means that
\beq\label{eq:Signs}
	|H(Q_i)+H(\bar D_j)-H(\Phi_1)| > |H(Q_i)+H(\bar D_j)-H(\Phi_2)|.
\eeq
Since $H(\Phi_1) > H(\Phi_2)$, this is only possible if the expression inside the absolute value on the left hand side is negative. The expression on the right hand side can be either positive or negative. It turns out that one scenario is not viable, while the other is very implausible.

\subsubsection{Wrong CKM Pattern}
If both expressions in Eq.~(\ref{eq:Signs}) are negative, then it turns out that the hierarchy between diagonal and off-diagonal entries in $Y^d$ is broken and it becomes impossible to get a CKM matrix that is close to the unit matrix, as in the following example.
\beqa\label{eq:chargei}
&& Q_{1,2,3}(3,2,0),\ \ \ \bar U_{1,2,3}(4,1,0),\ \ \ \bar D_{1,2,3}(-5,-6,-6),\no\\
&& L_{1,2,3}(1,1,1),\ \ \ \bar E_{1,2,3}(-9,-6,-4),\no\\
&&\Phi_{1,2}(2,0).
\eeqa
\beqa\label{eq:masmati}
&&Y^u_2\sim\begin{pmatrix}
	\epsilon_H^{7}&\epsilon_H^{4}&\epsilon_H^3\\
	\epsilon_H^{6}&\epsilon_H^{3}&\epsilon_H^2\\
	\epsilon_H^{4}&\epsilon_H^{1}&\epsilon_H^0\\
\end{pmatrix},\ \ \
(Y^u_1)_{ij}\sim\epsilon_H^{2}(Y^u_2)_{ij},\no\\
&&Y^d_2\sim
\begin{pmatrix}
	\epsilon_H^{2}&\epsilon_H^{3}&\epsilon_H^{3}\\
	\epsilon_H^{3}&\epsilon_H^{4}&\epsilon_H^{4}\\
	\epsilon_H^{5}&\epsilon_H^{6}&\epsilon_H^{6}\\
\end{pmatrix},\ \ \
(Y^d_1)_{ij}\sim\epsilon_H^{2}(Y^d_2)_{ij},\no\\
&&Y^e_2\sim
\begin{pmatrix}
	\epsilon_H^{8}&\epsilon_H^{5}&\epsilon_H^{3}\\
	\epsilon_H^{8}&\epsilon_H^{5}&\epsilon_H^{3}\\
	\epsilon_H^{8}&\epsilon_H^{5}&\epsilon_H^{3}\\
\end{pmatrix},\ \ \
(Y^e_1)_{ij}\sim \epsilon_H^2 (Y^e_2)_{ij}.
\eeqa
The resulting CKM pattern, in the standard convention where the quarks are ordered by masses, is not close to the unit matrix, but rather to a permutation of it,
\beq
V_{\rm CKM} = V_{uL}V_{dL}^\dagger\sim \begin{pmatrix}
	\epsilon_H^{3} & \epsilon_H  & 1\\
	\epsilon_H^2   & 1           & \epsilon_H\\
	1              &\epsilon_H^2 &\epsilon_H^{3}\\
\end{pmatrix}.
\eeq
%
%

\subsubsection{Large FN-Charge Difference}
If the expression on the right hand side of Eq.~(\ref{eq:Signs}) is positive, we have
\beq
	-H(Q_i)-H(\bar D_j)+H(\Phi_1) &>& H(Q_i) + H(\bar D_j) - H(\Phi_2) \\ \no
	\frac{1}{2}(H(\Phi_1)+H(\Phi_2)) & > & H(Q_i)+H(\bar D_j).
\eeq
Since this is true for all $i,j$, and we want to achieve $H(Q_1)+H(\bar D_1)-H(\Phi_2)=6(\log 0.2/\log \epsilon_H)$, we get a lower bound on the possible charge difference,
\beq
	H(\Phi_1)-H(\Phi_2) > 12\left(\frac{\log 0.2}{\log \epsilon_H}\right).
\eeq
Hence, a viable model can be built, but only at the cost of very large charge differences. An example, For Type I:
\beqa\label{eq:chargei}
&&Q_{1,2,3}(6,5,3),\ \ \ \bar U_{1,2,3}(4,1,0),\ \ \ \bar D_{1,2,3}(-3,-4,-4),\no\\
&&L_{1,2,3}(1,1,1),\ \ \ \bar E_{1,2,3}(-12,-9,-7),\no\\
&&\Phi_{1,2}(+10,-3).
\eeqa
The parametric suppression of the matrix elements is given by
\beqa\label{eq:masmati}
&&Y^u_2\sim\begin{pmatrix}
	\epsilon_H^{7}&\epsilon_H^{4}&\epsilon_H^3\\
	\epsilon_H^{6}&\epsilon_H^{3}&\epsilon_H^2\\
	\epsilon_H^{4}&\epsilon_H^{1}&\epsilon_H^0\\
\end{pmatrix},\ \ \
(Y^u_1)_{ij}\sim\epsilon_H^{13}(Y^u_2)_{ij},\no\\
&&Y^d_2\sim
\begin{pmatrix}
	\epsilon_H^{6}&\epsilon_H^{5}&\epsilon_H^{5}\\
	\epsilon_H^{5}&\epsilon_H^{4}&\epsilon_H^{4}\\
	\epsilon_H^{3}&\epsilon_H^{2}&\epsilon_H^{2}\\
\end{pmatrix},\ \ \
Y^d_1\sim
\begin{pmatrix}
	\epsilon_H^{7}&\epsilon_H^{8}&\epsilon_H^{8}\\
	\epsilon_H^{8}&\epsilon_H^{9}&\epsilon_H^{9}\\
	\epsilon_H^{10}&\epsilon_H^{11}&\epsilon_H^{11}\\
\end{pmatrix},\no\\
&&Y^e_2\sim
\begin{pmatrix}
	\epsilon_H^{8}&\epsilon_H^{5}&\epsilon_H^{3}\\
	\epsilon_H^{8}&\epsilon_H^{5}&\epsilon_H^{3}\\
	\epsilon_H^{8}&\epsilon_H^{5}&\epsilon_H^{3}\\
\end{pmatrix},\ \ \
(Y^e_1)_{ij}\sim \epsilon_H^{13} (Y^e_2)_{ij}.
\eeqa
We note that the parametric suppression of down $Y_1^d$ compared to $Y_2^d$ is no longer universal.

A choice with opposite signs for $H(L_i)+H(\bar E_j)$ in the lepton sector gives a model of Type III with similar features.

Given that we introduce the FN mechanism to explain the smallness of, among others, $|V_{us}|\sim0.2$, or $m_\mu/m_\tau\sim0.06$, and that we do it here by introducing charge ratios as small as $1/12\sim0.08$, it seems to us that this class of models, even if phenomenologically viable, misses the main goal of introducing the FN mechanism.



\end{document}